# Virtual Placebo: a scoping review of control conditions in randomized controlled trials of psychological immersive virtual reality interventions

**Madis Vasser[1], Mary-Ann Kubre[1], Kadi Tulver [2] and Jaan Aru[1]**

[1] Institute of Computer Science, University of Tartu, Estonia
[2] Institute of Psychology, University of Leiden, the Netherlands
Corresponding author: Madis Vasser, madis.vasser@ut.ee

**Abstract**

Research on immersive virtual reality (IVR) as a therapeutic tool is growing rapidly, with randomized controlled trials (RCTs) now commonplace. Yet the control conditions in many of these studies that claim strong specific IVR effects vary from usual care to active immersive experiences, making it difficult to draw clinically meaningful conclusions. We conducted a scoping review of recent literature on psychological IVR interventions to assess the main forms of control conditions used and their relations to study outcomes. In this paper, we make the case that most of the recent RCTs studying the effects of specific IVR interventions compare active immersive VR to usual care with no multimedia component. This can lead to strong yet potentially misleading study outcomes and can potentially inflate the IVR interventions' effect. Namely, the lack of an appropriate immersive control condition may risk misattributing general effects of VR devices to the specific content of the intervention. In this review, we discuss the implications of these design choices and propose strategies for including a suitable "virtual placebo" condition to improve the methodological rigor and interpretability of results.

## 1. Introduction

The term "virtual reality" (VR) is used in the literature to describe experiences employing tracked immersive head-mounted displays (either standalone devices or ones that are connected to a computer) and hand controllers, but it is sometimes also used for video games played on a conventional 2D screen. In this review, we adopt the technical definition of immersive VR (IVR) offered by Rizzo & Koenig (2017, p. 878): "Immersive VR can be produced by the integration of computers, head-mounted displays (HMDs), body-tracking sensors, specialized interface devices, and 3D graphics". So far, clinical immersive virtual reality (clinical IVR) has been used in various therapeutic settings with positive outcomes reported (see Emmelkamp & Meyerbröker, 2021; Meshkat et al, 2024; Quan et al, 2024). While IVR elicits well-established place (the subjective feeling of being present in the virtual environment) and plausibility (virtual events seeming credible to the user and prompting a realistic response to the virtual reality) illusions (Slater, 2009), the underlying mechanisms of the therapeutic benefits of IVR are yet to be identified and more research on these topics is still needed (Rizzo & Koenig, 2017).

The therapeutic potential of IVR has been primarily attributed to its ability to provide patients with a safe, controllable, and immersive environment in which to engage with stressful or otherwise unpleasant situations under a therapist's guidance (Gregg & Tarrier, 2007; Emmelkamp & Meyerbröker, 2021). This is most evident in virtual reality exposure therapy (VRET), where patients gradually confront their avoided stimuli or situations within a realistic yet secure virtual setting, making it an especially pertinent tool for the treatment of anxiety disorders and trauma and/or stressor related disorders (e.g., specific phobias, PTSD), but also substance use and eating disorders (for a review, see Emmelkamp & Meyerbröker, 2021). Beyond phobias and anxiety disorders, IVR environments have also been used to train social and cognitive skills in conditions such as attention deficit hyperactive disorder, autism spectrum disorder, and psychosis (Emmelkamp & Meyerbröker, 2021), or to facilitate emotional regulation and relaxation through guided imagery and interactive visualization (e.g., Chuan et al., 2023; Rutkowski et al., 2023).

However, much like other psychological interventions, IVR's benefits may not arise solely from its specific content or mechanisms. Factors such as patients' expectations (e.g., receiving a high-tech intervention), emotional engagement (e.g., interacting with the therapist), and the contextual cues (e.g., donning the VR goggles) surrounding the IVR experience, which are the core components that are often attributed to placebo effects (Price et al., 2008), may also contribute to therapeutic outcomes. Specifically, the proposed effects of novelty and gamification have been noted as a potential confounding variable when using technological interventions, including IVR and AI (Kosch et al., 2023; Miguel-Alonso et al., 2024). Researchers must be able to distinguish between improvements driven by the specific therapeutic content (e.g., exposure to spiders) and those stemming from non-specific VR effects (e.g., immersion, distraction, or positive expectations), making the design of appropriate control conditions in IVR trials particularly challenging.

Clinical IVR is often used in randomized controlled trials (RCTs), which is a gold standard approach in clinical research for assessing the efficacy of an intervention. A key feature of such trials is the inclusion of a control condition against which the effect of a specific intervention can be compared. For instance, a placebo control group may be used in which the intervention closely resembles the treatment group but lacks the "active ingredient" (i.e., a sham condition). A classic example is the sugar pill in pharmaceutical studies. This approach is most commonly used in double-blind RCTs, which are designed to conceal the nature of the treatment from both study participants and researchers conducting the experiment, thereby helping to minimize the likelihood of unintentionally skewing the study results due to the participants' and/or researchers' (unconscious) expectations or biases.

To further facilitate research in the field, an international working group called the Virtual Reality Clinical Outcomes Research Experts (VR-CORE) has outlined a hierarchy of control conditions for advanced phase III clinical randomized controlled VR studies that evaluate efficacy against a control condition (Birckhead et al, 2019). These range from usual care (i.e., without any active intervention) all the way to immersive and active (i.e., interactive) experiences within a VR HMD. This differs from phase I studies, which focus more on content development, and phase II trials that conduct early testing with a focus on feasibility, acceptability, tolerability, and initial clinical efficacy (Birckhead et al, 2019). Establishing the Rizzo and Koenig (2017) highlighted that while IVR was emerging as a promising tool for professionals, clear guidelines for its use in research were still lacking. Despite these observations, no formal guidelines have yet been established as of February 2026, leaving clinical IVR research without a universally accepted gold standard for RCT design.

Given this gap, we conducted a scoping review to obtain an overview of previous and current practices in clinical IVR RCT control conditions. We focused specifically on IVR in psychological interventions to capture the present state of the rapidly developing and changing field and to identify emerging trends. This narrower focus exclusively on immersive VR instead of all types of VR (e.g. desktop- and smartphone-based approaches) also helped ensure that the included studies are more comparable, as both VR hardware and software have – and still are – developed at a rapid pace. In addition to an overview of the IVR landscape, we also aim to highlight methodological challenges that might have to be addressed in future studies (Page et al, 2021). This can help advance the field and support the development of reproducible, high-quality, and clinically relevant studies.

## 2. Literature review

The aim of a literature review is to assemble, arrange, and assess the existing literature in a given domain: using IVR as a therapeutic tool in cases of various psychological conditions, symptoms, or quality of life. For our review, we followed the PRISMA 2020 scoping review guidelines (Tricco et al, 2018; see Appendix 1). We defined our literature review question as: "What type of interventions and control conditions have been used in randomized controlled immersive clinical VR trials that study the effect of specific VR content on various psychological conditions, symptoms or quality of life?". We defined psychological conditions and symptoms as those that can be found in manuals such as the International Classification of Diseases 11th Revision (World Health Organization, 2019) or the Diagnostic and Statistical Manual of Mental Disorders Fifth Edition (American Psychiatric Association, 2022) and that were evaluated using the appropriate questionnaire(s). Such a review question formulation covers not only psychological IVR interventions, but also clinical IVR interventions with psychological outcomes and broader clinical IVR studies with quality of life measures as a primary or secondary outcome. Since our focus was specifically on applications of IVR, a small subgroup of the larger VR research field, we reasoned that including a broader set of interventions was justified to reach a bigger sample of initial articles. We conducted our search in January 2026 among articles listed in the databases of PubMed, Web of Science, Scopus, and PsycINFO. We searched for "immersive virtual reality" and "randomized controlled trial" in the title field of journal articles. We also used several search keywords in order to exclude titles related to non-immersive devices (not "non-immersive"), reviews (not "review" or "meta-analysis"), pre-registrations (not "protocol"), physiological interventions (not "training", "physiotherapy", "parkinson", "balance" or "stroke") or non-specific effects of IVR (not "distraction"). We set no restrictions on the publishing date. We decided to exclude review articles because early scoping revealed that these tend to gloss over the finer details regarding the control conditions of the included studies, making it difficult to classify and assess the true nature of the control conditions. Also, some review articles specifically exclude studies that use IVR as a control condition (e.g., Ng et al, 2019), which is not in line with our scope. We also omitted pre-registrations of research protocols and trials as we were interested in evaluating the general direction of the results of the studies (significant or non-significant differences between groups). An example search syntax from the PsycINFO database was the following (PZ stands for publication type, TI for title): PZ ("journal article") AND TI ("immersive virtual reality" AND "randomized controlled trial") NOT TI (non-immersive OR review OR protocol OR distraction OR training OR physiotherapy OR parkinson OR balance OR stroke OR meta-analysis). See Appendix 2 for all the full search queries.

Relying on title-field searches for immersive VR may have excluded relevant studies that used alternative terminology and biased the results towards studies that use high-end VR hardware. However, expanding the search to abstract and keyword fields would have also resulted in a very high number of (desktop- and smartphone-based) studies in our initial sample and we deemed this counter-productive from a scoping review standpoint as including a wide variety of devices makes comparing the studies more challenging.

### *2.1. Literature review results*

The initial database search resulted in a total of 135 journal articles written in English (no articles in other languages were found). After removing 88 duplicates, the article count was 47. Since we were specifically interested in RCTs that study IVR (rather than screen-based interventions) and content-specific effects (rather than non-specific effects of IVR-based distractions in general) on psychological conditions or symptoms (rather than medical personnel training, or balance and motor training), further filtering was undertaken. By assessing the title, abstract and full text, we further excluded articles that:

1) did not exclusively use an immersive wide field-of-view HMD (e.g., used smartphone-based VR devices) (11 articles removed);

2) did not measure a change in a psychological condition and/or symptom or quality of life as the primary or secondary outcome of the intervention, or did not report results, e.g., pre-registrations (14 articles removed).

We defined high-fidelity HMDs as those either connected to a computer via a cable (e.g., Oculus Rift, HTC Vive) or as standalone devices with dedicated processors (e.g., Oculus Go, Meta Quest, PICO) that offer a wide field of view. We made this decision so that the included studies would be as similar as possible in terms of hardware and thus comparable to each other.

The final number of articles included was 22 (see Figure 1). All included studies can be found in Table 1. We did not filter articles based on the perceived clinical phase, as being a RCT already suggests that the study can be considered phase III.

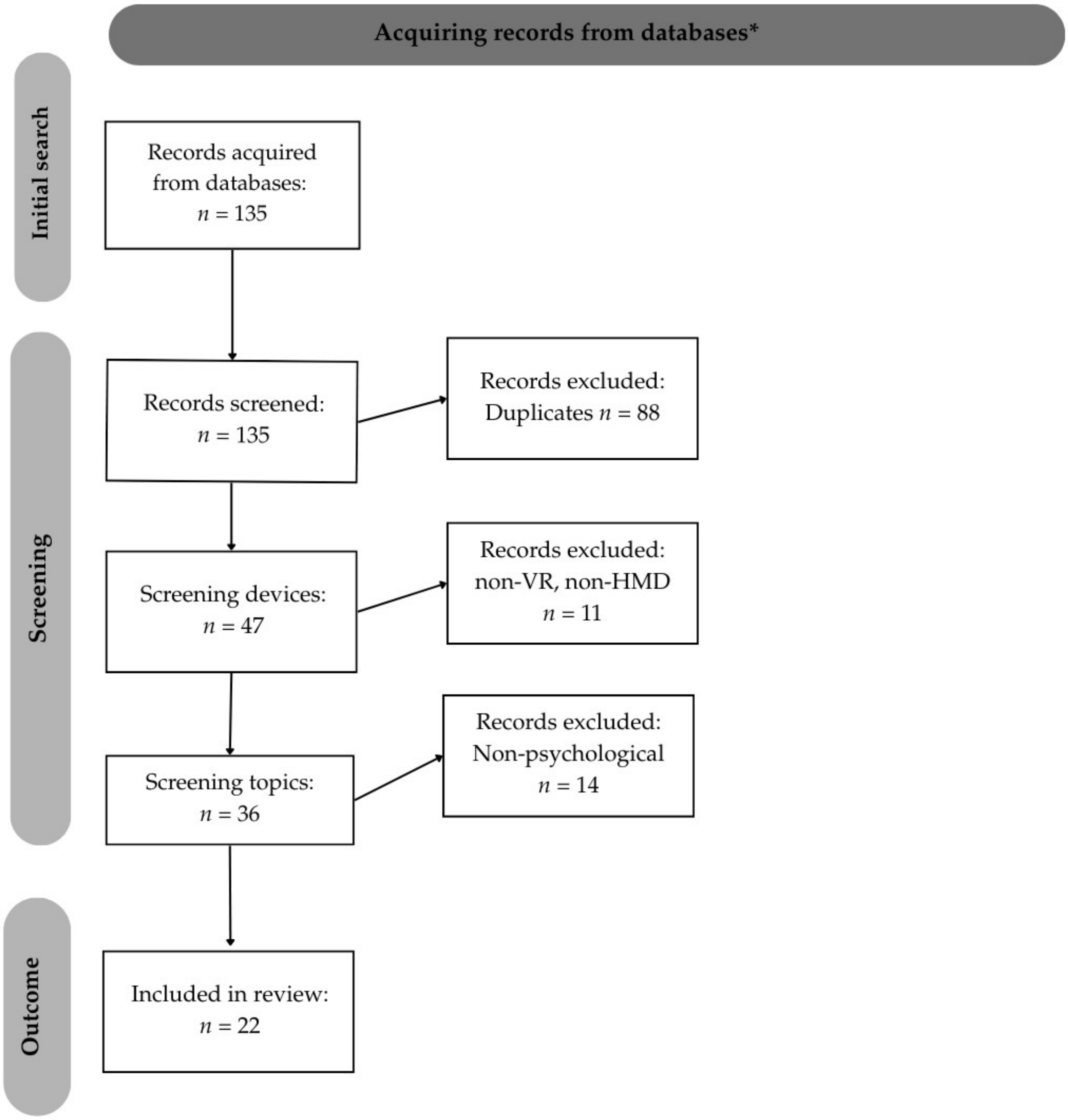

**Figure 1.** Flowchart of the review process. * PubMed, Web of Science, Scopus and PsycINFO.

We continued our analysis with a total of 22 articles that fit our inclusion criteria. Two of the authors (MV and MAK) assessed the data from the publications by categorizing the studies by their control conditions according to VR-CORE recommendations (Birckhead et al, 2019). The proposed five levels for the control conditions by Birckhead and colleagues (2019) are the following, with our own short-hand codes used in the current review in parenthesis:

1) usual care or waiting list without any active intervention (No intervention);
2) passive visualizations on a two-dimensional screen (Passive non-VR), and as our own addition, other non-screen-based activities that differ from usual care (Passive or active non-VR);
3) non-immersive visualizations within an HMD (Non-immersive VR);
4) immersive but passive experiences with no task within an HMD (Passive IVR);
5) immersive and active experiences with a task within an HMD (Active IVR).

The initial inter-rater reliability for classifying the experimental and control conditions of all the studies included in the final review was 95**%**, with any remaining discrepancies resolved by discussion. One discrepancy that emerged from the VR CORE guidelines was what exactly constituted "usual care", unless the authors explicitly used said – or a similar term – to describe the control condition. Thus, we broadened our "no intervention" category to accommodate any conditions that did not involve a multimedia component (e.g., text-based therapy, waiting list, usual care).

The only other point of discrepancy was regarding what constituted passive *versus* active immersive VR. After discussing, we agreed among ourselves that conditions where the stimulus itself is passive (e.g., watching an immersive non-interactive video) but the participant must execute some kind of an IVR experience related task (e.g., answer some questions regarding the video, execute a guided exercise, etc.) either during or after the VR experience, should be marked as active IVR, as the participant is not passive and has an ongoing task. As a concrete example, participants were watching 360-degree videos but also had to follow the audio tour guide and conduct a savoring exercise after the videos (Pancini et al, 2025). Conversely, we categorized conditions where participants viewed visualizations but had no specific VR experience related task as passive immersive. We did not thoroughly assess the outcomes or the overall risk of bias in the reported results of the included studies. Risk of bias assessments are more applicable to systematic reviews of interventions rather than our scoping review on methods. Our focus was not on whether the included studies were properly conducted *per se*, but rather on the type of control conditions and on whether there were any reported statistically significant differences between the conditions. However, IVR studies with weaker controls may also differ in sample size, statistical power, intervention quality, publication year, or reporting practices.

The specific domains of the studies were cognitive therapy ($n$ = 16), pain management ($n$ = 4), and rehabilitation ($n$ = 2). Interventions ranged from active IVR ($n$ = 14) to passive IVR ($n$ = 8). The control conditions were as follows: no intervention ($n$ = 14), passive non-VR ($n$ = 4), passive IVR ($n$ = 3), and active non-VR
($n$ = 1).

**Table 1.** Overview of the articles included in the review.

| Authors, year | Domain | Relevant measurements | Control condition | Intervention | Type of HMD | Significant differences between groups |
|---|---|---|---|---|---|---|
| Abd El Fatah et al, 2024 | Cognitive therapy | Mini-Mental State Examination (MMSE); Reminiscence Psychological Well-Being Scale (RPWBS) | Passive non-VR (Reminiscence therapy) | Active IVR (VR Wander, custom content) | Oculus Quest | No; Both groups significantly improved |
| Bai et al, 2025 | Cognitive therapy | Modified Yale Perioperative Anxiety Scale (m-YPAS) | Passive IVR (custom games) | Active IVR (custom games) | Pico Neo3 | No; Both groups significantly improved |
| Burrai et al, 2023 | Cognitive therapy | State-Trait Anxiety Inventory-Y (STAI-Y1) | Passive non-VR (Narrative medicine) | Passive IVR (360 video) | Oculus Quest 2 | Yes; VR group significantly improved compared to control |
| Çetin et al, 2025 | Pain management | The Short Form Health Questionnaire-36 (SF-36) | No intervention | Passive IVR (Ocean Rift) | Oculus Go | No; No significant differences between groups |
| Chang et al, 2022 | Pain management | McMurtry Children's Fear Scale (CFS) | No intervention | Passive IVR (SILVER) | Oculus Quest | Yes; VR group significantly improved compared to control |
| Cieślik et al, 2023 | Cognitive therapy | Geriatric depression scale (GDS); Hospital Anxiety and Depression Scale (HADS) | No intervention | Active IVR (VRTierOne) | HTC Vive | Yes; VR group significantly improved compared to control |
| Freeman et al, 2018 | Cognitive therapy | Heights Interpretation Questionnaire (HIQ) | No intervention | Active IVR (Now I Can Do Heights) | HTC Vive | Yes; VR group significantly improved compared to control |
| Jiménez et al, 2025 | Cognitive therapy | Edinburgh Postnatal Depression Scale (EPDS); State-Trait Anxiety Inventory | No intervention | Active IVR (Custom content) | Oculus Go | VR group significantly improved compared to control. |

| Authors, year | Domain | Relevant measurements | Control condition | Intervention | Type of HMD | Significant differences between groups |
|---|---|---|---|---|---|---|
| Kamada et al, 2025 | Cognitive therapy | Intraoperative and postoperative anxiety, visual analog scale | No intervention | Active IVR (Therapeia VR) | Pico G2 | Yes; VR group significantly improved compared to control. |
| Kwan et al, 2026 | Cognitive therapy | World Health Organisation Five Well-being Index; Patient Health Questionnaire; EuroQol 5-Dimensions instrument | No intervention | Passive IVR (360 video) | Oculus Quest 2 | Yes; VR group significantly improved compared to control. |
| Liu et al, 2025 | Cognitive therapy | Brunel Mood Scale; Beck Depression Inventory | No intervention | Active IVR (VirZoom VR) | PlayStation VR | Yes; VR group significantly improved compared to control. |
| Mazurek et al, 2023 | Rehabilitation | Hospital Anxiety and Depression Scale (HADS) | No intervention | Active IVR (VRTierOne) | HTC Vive Pro | Yes; VR group significantly improved compared to control |
| Niki et al, 2025 | Cognitive therapy | Edmonton Symptom Rating System (ESAS-r-J);State-trait anxiety inventory-JYZ (STAI) Y-1 | No intervention | Passive IVR (Commercial games) | Oculus Go | Yes; VR group significantly improved compared to control. |
| Pancini et al, 2025 | Cognitive therapy | Italian Mental Health Continuum Short Form (MHC-SF) | Passive non-VR (music listening | Active IVR (360 video and related tasks) | Oculus Quest 2 | Yes; VR group significantly improved compared to control. |
| Parmak et al, 2025 | Cognitive therapy | The World Health Organization Quality of Life Instrument—Older Adults Module (WHOQOL-OLD) | No intervention | Active IVR (FIT-XR) | Meta Quest 2 | Yes; VR group significantly improved compared to control. |

| Authors, year | Domain | Relevant measurements | Control condition | Intervention | Type of HMD | Significant differences between groups |
|---|---|---|---|---|---|---|
| Perdue et al, 2023 | Pain management | Anxiety visual analog scale | Passive non-VR (wearing HMD) | Passive IVR (Ocean Rift) | Oculus Go | No;<br>No significant differences between groups |
| Prahm et al, 2025 | Rehabilitation | Short Form Health Survey (SF-36) | Passive IVR (360 videos) | Active IVR (StableHandVR) | Meta Quest 2/3 | No;<br>Both groups significantly improved, with some significant differences between groups. |
| Rutkowski et al, 2021 | Cognitive therapy | Hospital Anxiety and Depression Scale (HADS) | No intervention | Active IVR (VRTierOne) | HTC Vive | Yes;<br>VR group significantly improved compared to control |
| Selvaraj et al, 2025 | Pain management | Cambridge Depersonalization Scale (CDS) | Passive IVR (2D animations) | Passive IVR (Custom content) | Oculus Quest 2 | Yes;<br>VR group significantly improved compared to control. |
| Shaw & Lubetzky, 2021 | Cognitive therapy | Pediatric Anxiety Short Form 8a (PASF) | Active non-VR (Dodgeball) | Active IVR (Custom game) | Oculus Rift | No;<br>Both groups significantly improved |
| Szczepańska-Gieracha et al, 2021 | Cognitive therapy | Hospital Anxiety and Depression Scale (HADS) | No intervention | Active IVR (VRTierOne) | HTC Vive | Yes;<br>VR group significantly improved compared to control |
| Turgut et al, 2024 | Cognitive therapy | Children's State Anxiety Scale | No intervention | Passive IVR (360 video) | Oculus Go | Yes;<br>VR group significantly improved compared to control |

NB “No intervention” includes both usual care and waiting list controls.

*2.2 Inferential statistics*

Given the small sample size, wide confidence intervals, and included studies being highly heterogeneous (in terms of study quality, risk of bias, population, clinical settings, intervention aims, protocol, outcomes, effect size, duration and intensity of IVR), we conducted only a binary vote-counting of statistical significance across studies. Our results should therefore be taken as exploratory and hypothesis-generating for further systematic research.

We compared the prevalence of finding a statistically significant result between studies using no intervention control groups (Group 1, $n$ = 14) and articles employing some sort of control condition (Group 2, $n$ = 8). In Group 1, 92.91% (13 out of 14) of studies found a statistical effect, compared to only 37.5% (3 out of 8) in Group 2, where usually the scores of both control and intervention groups improved (see Figure 2). This difference between the groups was statistically significant (two-sided Fisher's exact $p$ = .011). The effect size was large, with a Haldane–Anscombe corrected odds ratio of 14.14 (95% CI [1.64, 121.96], applied to account for small cell counts. While the confidence interval for the odds ratio is wide because of the small sample of studies with intervention controls, this result hints that regardless of content, an IVR experience in itself could lead to overall clinical improvements. Also, it suggests a high probability of finding a statistically significant result in an IVR study with no intervention control groups compared to an IVR study with an intervention control group.

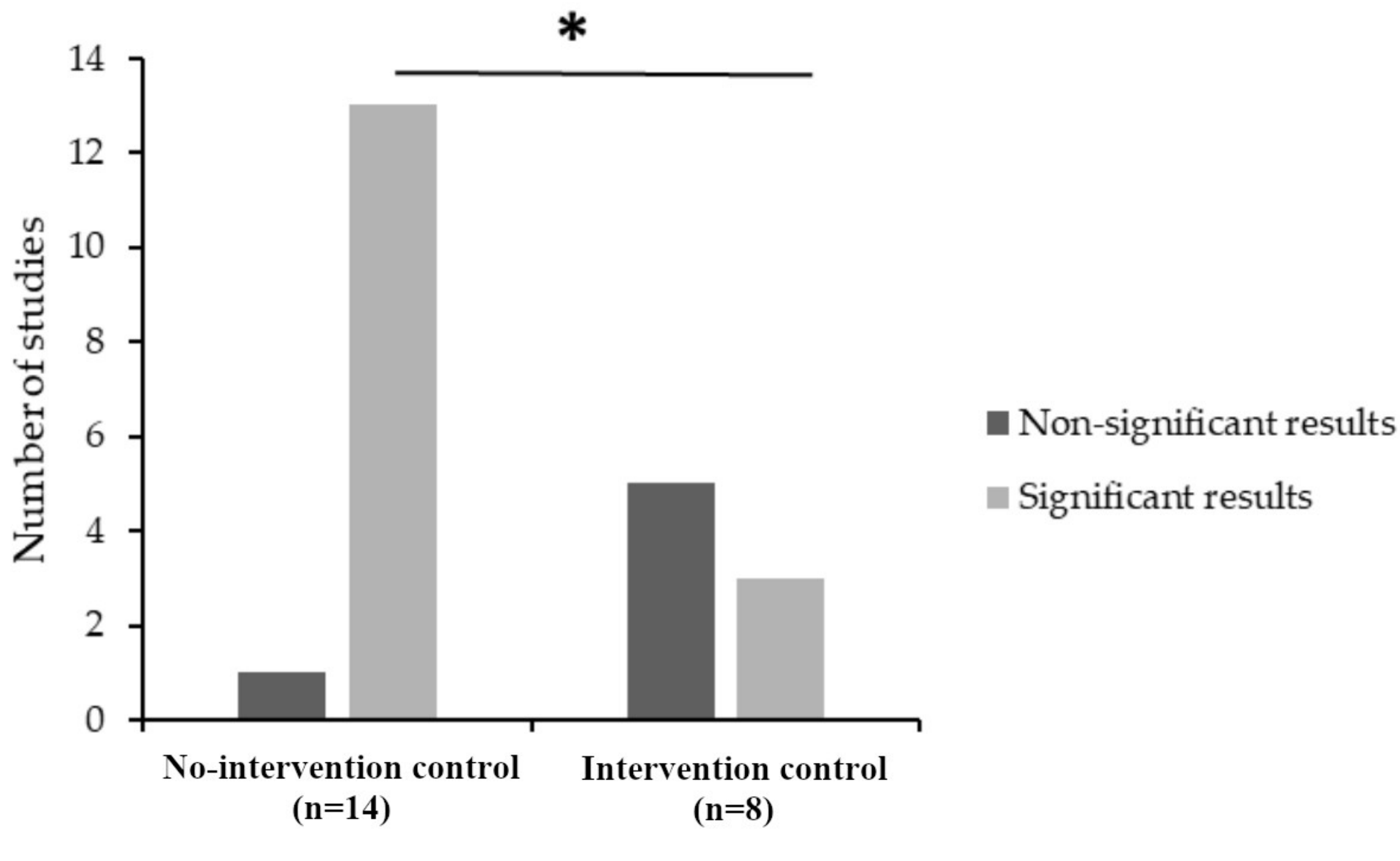

**Figure 2.** The number of included studies (y-axis) categorized by control condition (x-axis) and outcome (light gray – significant results, dark gray – non-significant results). The prevalence of statistically significant results was compared between studies using no-intervention control groups (n = 14) and those employing some sort of intervention control conditions (n = 8).

## 3. Discussion

These results raise the important question of what could be driving these potential non-specific benefits of IVR. Previous work has suggested that VR technology itself may elicit a strong placebo effect in study participants (Rizzo & Koenig, 2017). As IVR is still relatively unfamiliar to many participants, even the use of this novel hardware may enhance engagement and positive expectations, regardless of the specifically designed software or content. Although the mechanisms underlying the beneficial effects of IVR are still debated, there is some evidence to suggest that the higher levels of engagement in immersive technologies can significantly improve patients' emotional states and well-being (Buche et al., 2025). Beyond mechanisms that can influence participants, some bias may also emerge at the level of the person designing or conducting the experiments. This is known in the clinical literature as "researcher's therapy allegiance", where *a priori* devotion to a treatment mode might skew outcomes (Luborsky et al, 1999). Based on this theory, high enthusiasm for a novel technological solution could inadvertently influence study design, implementation, or interpretation, for instance, by the IVR researcher unintentionally communicating higher expectations or selecting a less effective competing treatment for comparison with their favored treatment. While this approach may not be prevalent in the IVR research field per se, it is something to be cautious about when designing any randomized controlled studies.

There is no one "perfect" control condition for all VR treatments, as the suitable control condition depends on things such as the patient population, clinical setting and the proposed mechanism of the studied intervention (Birckhead et al., 2019). Additionally, lack of access to specific solutions may also play a part in choosing a control condition. Yet we believe that comparing an IVR intervention to no intervention might not be optimal for drawing clinical conclusions about specific IVR programs for a number of reasons.

Firstly, the inherent nature of HMD-based IVR provides a clear separation between reality and VR, dubbed the device-gap (Slater et al, 2020). This makes it obvious for both the participants and experimenters whether they are currently involved in the active intervention condition (by wearing VR goggles) which can bias their behavior inadvertently. This can then skew the outcome of the intervention. Secondly, some non-immersive control conditions may seem less credible as clinical interventions to study participants, which have also been shown to influence study outcomes when control conditions are compromised (Mohr et al, 2009). On the other hand, the awareness of being in a control or placebo group can itself elicit clinical effects in (some of) the participants (Fendel et al, 2025). Therefore, control conditions can have various effects on participants. Issues can also arise when non-immersive conditions are used as the control condition for IVR, as the former differ significantly from the IVR experience. However, using commercial VR games as controls (suggested by Monti and Aglioti, 2018) can have its own downsides, as the games might be too engaging compared to the potentially plainer experimental software and thus skew the data in another direction.

Thus, it is difficult to choose the ideal IVR control condition that could effectively mask the study group the participant belongs to. Guidelines for RCTs state that in placebo-controlled trials, effort should be put into producing a placebo that is as similar to the intervention as possible (Spieth et al., 2016). Granted, this is not an easy task and often no relevant guidelines exist, so researchers are called to "bring creativity and ingenuity to the design of control conditions, based on the study questions, feasibility issues, threats to internal validity, and other needs of the study" (Mohr et al., 2009, p. 280). Adequate control conditions are especially important for later stage (phase III) clinical trials that are the final step before potential wider rollout to practitioners. If the study hypotheses concern mechanisms linked to specific IVR content (e.g., a snowy scene to elicit a cooling effect to reduce fever), the use of nonspecific controls (i.e., VR in general) may be both appropriate and necessary (Mohr et al, 2009). Otherwise, there remains a risk of misattributing any beneficial effects to the wrong component of the intervention, such as deploying a very specific and possibly costly "Winter IVR" software, while a cheaper nonspecific IVR experience (a random immersive video) would work equally well. To assist researchers in picking the most suitable control condition, we will propose some potential solutions – instead of using usual care or no intervention as controls – in the next section.

### *3.1. Potential solutions for virtual placebos*

The VR-CORE experts (Birckhead et al, 2019) recommend balancing the experimental and placebo conditions by exposing participants in both study arms to HMDs, but varying the content being experienced, e.g., immersive *versus* non-immersive or active *versus* passive content to demonstrate whether interventions are effective. Compared to pharmaceutical studies, selecting the most appropriate placebo conditions in experiments investigating the effectiveness of behavioral interventions (under which IVR falls) is less clear (Hecksteden et al., 2018), and concrete recommendations for designing specifically IVR control conditions are currently lacking. Ideally, any new clinical IVR intervention should be controlled against the best currently available IVR treatment. It might, however, be unclear as to what that is – or whether such an intervention already exists – which makes the process of choosing the most suitable control condition challenging. This is because IVR interventions are not as rigorously managed as other clinical interventions (e.g., pharmaceutical drugs). We build upon the suggestion by VR-CORE experts and propose developing IVR control experiences that closely resemble the experimental condition in both hardware and software domains, with the main difference being the specific therapeutic mechanism under study when possible. We propose a set of design recommendations for constructing such 'virtual placebo' controls, summarized in Table 2. For example, for a snowy IVR intervention scene that is hypothesized to lower fever via inducing a cooling effect, a suitable control would be a non-snowy IVR environment with the same level of interactivity (i.e., as passive or active as the intervention). Such a control condition could be more appropriate as the suspected therapeutic mechanisms of the intervention are directly compared to another VR intervention. Additionally, using HMDs for all conditions can help blind the participants and researchers to the condition and can thus help further evaluate the effect of the content of the IVR experience as all participants will be required to wear the HMD and either execute similar tasks or passively observe the VR experience. Moreover, recent systematic reviews have found slightly lower dropout rates in VR groups than in non-VR controls (García-Muñoz, Cortés-Vega, & Martínez-Miranda, 2025), further supporting the use of immersive controls to balance participant engagement across groups.

**Table 2**. Recommendations for incorporating virtual placebo as a control condition when designing clinical IVR RCT studies.

| Design Aspect | Intervention vs. Control | Recommendation | Rationale |
|---|---|---|---|
| Hardware | Equivalent | Use the same HMD and controllers for both groups. | Controls for hardware-related expectancy effects and aims to maintain comparable levels of immersion and sensory experience. |
| Exposure Duration and Frequency | Equivalent | Keep exposure time, session length, and frequency identical across groups. | Prevents unequal "dose" effects between intervention and control. |
| Level of Activity | Equivalent | Match the degree of interactivity (e.g., both passive observation or both active engagement). | Balances cognitive load and engagement across conditions. |
| Audiovisual Fidelity | Comparable | Aim for similar graphical quality, level of detail, and sound environment (unless these are the independent variables under study). | Maintains ecological validity and immersion while isolating mechanism-related differences. |
| Therapeutic Mechanism | Distinct | Remove or alter the hypothesized active component. | Isolates the specific therapeutic mechanism under investigation. |

Ideally, a virtual placebo should strive for complex and realistic scenes with high graphical fidelity, which would increase the ecological validity of the experience (Ramirez, 2019) and ensure a low drop-out rate (Rizzo & Koenig, 2017; Vasser & Aru, 2020). This assumes that the IVR intervention condition also shares similar properties to the control condition. The control condition must ideally strike a balance between not being too boring nor too interesting. A perfect generic and nonspecific virtual experience would fit a range of very different immersive interventions.

In practice, implementing and validating a virtual placebo experience could be as follows. First, researchers design and develop or acquire the IVR intervention condition (e.g. a virtual psychedelic tunnel where participants are tasked with catching flying angels with their hands, to study the emotional effects of mystical type experiences). Then, the virtual placebo conditions must be made or found, comprising a very similar overall look and feel, e.g. a virtual mundane-looking pedestrian tunnel where participants must catch flying debris. Thus, the participants are exposed to equivalent geometries and movement, with the key difference being the content of the visuals and perhaps audio. To the outside observer, the participants' behavior would look identical. The resulting data would possibly show some emotional effects in both conditions – a fully inert virtual placebo is realistically not achievable due to strong nonspecific effects in immersive technologies. The key question would be if the intervention condition significantly differs from the virtual placebo. How to ensure that a virtual placebo is neutral with respect to the targeted therapeutic mechanism, i.e. inert like a pharmaceutical sugar pill, remains case specific. Such a virtual placebo would clearly differ from a sham intervention (e.g. wearing a non-functional HMD), attention control (e.g. watching a 2D video), and an active comparator (e.g. another IVR experience with different specific visuals and tasks).

A virtual placebo can require additional time and financial means for finding, developing or licensing a comparable immersive control condition that is suitable for the specific intervention - resources that some researchers might not have readily available. One potential solution would be to create a library of virtual placebo IVR experiences of varying duration, immersion, and activity levels to try to cater to different experimental conditions. Rather than requiring every IVR researcher to create custom control software, these pre-existing placebo environments could be shared and reused across studies. Ideally, these control conditions should be customizable (e.g., in duration) to accommodate more researchers and their needs. To our knowledge, there is currently one freely available immersive IVR experience specifically developed as a general non-specific virtual placebo for clinical research (Routine Realms;[1]Siimon et al., 2023, preprint). The IVR program consists of mundane everyday environments such as a hotel room, a meeting room and a museum hall, and offers highly realistic scenes with either active or passive gameplay and automatic switching between different environments. While this tool has not yet been thoroughly validated, it could potentially offer some researchers a suitable alternative for an intervention with the same level of interactivity. We hope that more IVR software developers follow suit and create their own interpretations of a virtual placebo that others could use in their experiments of clinical IVR. Hopefully, this could shed some more light on how clinical IVR affects various (psychological) syndromes and/or diagnoses. Rigorous control design for digital health trials is not merely a technical requirement, but also an ethical imperative (Druart & Blease, 2025).

## 4. Limitations

Despite providing a brief overview of a specific topic, a scoping review also has its drawbacks. Accelerating the data extraction process may lead to missing some relevant information, and biases may be introduced due to shortened timeframes for literature search, article retrieval, and appraisal (Ganann, Ciliska, &

[1] https://nail.cs.ut.ee/routinerealms

Thomas, 2010). Still, studies have shown that overall, the conclusions between scoping reviews and systematic reviews tend to be congruent, thus reaching similar conclusions (Tricco et al, 2015). Due to our narrow research question, only 22 articles met the inclusion criteria, which is in the lower quartile of published peer-review scoping reviews (Tricco et al, 2016). Given the overall large volume of published VR research, our small sample size does not fully represent the broader research landscape but does provide an initial overview of control conditions used in IVR studies that have psychological conditions and/or quality of life as (one of) the outcome variables.

Future research should extend the research scope to investigate the effects in other domains, as well as with different hardware, too. We also did not have a pre-registered study protocol, which is a clear limitation that future work can overcome by building on our reported methodology.

While we categorized the intervention and control conditions according to VR-CORE recommendations (Birckhead et al, 2019), these instructions were quite broad, at times lacking the necessary specificity. Thus, some inherent subjectivity was introduced to our classification, although the inter-rater reliability was high. In addition, this review focused on methodological aspects of study design rather than on treatment outcomes or effect sizes. Our findings should therefore be interpreted as indicative of research practices rather than clinical efficacy.

Finally, our proposed "virtual placebo" is intended as a conceptual starting point for developing more standardized guidelines. As researchers begin to implement and refine this approach in practice, it can provide a foundation for establishing shared methodological standards in clinical IVR research.

## 5. Conclusion

We conducted a scoping methodological literature review of published randomized controlled clinical IVR studies with a focus on the used control conditions. We found that most of the reviewed studies used no intervention as the control condition for IVR, contrary to the calls for more rigorous research practices (Rizzo & Koenig, 2017) and guidelines from VR-CORE (Birckhead et al, 2019). Only 8 out of 22 studies (36%) in our sample used some sort of intervention control conditions. This could be caused by a few reasons – e.g., budgetary restrictions, researcher bias or common practice in the field to use no intervention controls. Such approaches harbor the risk of misattributing potential clinical effects to a specific IVR intervention, while simply using IVR in general might offer comparable therapeutic benefits. We propose using a virtual placebo – a control condition that matches the intervention in immersion and activity – to help minimize the non-specific effects related to IVR. Additionally, we propose creating a database of already existing IVR control conditions to increase accessibility to such tools. This is vital as using an appropriate IVR control condition would allow researchers to better investigate potential mechanisms of the therapeutic effects of clinical IVR interventions.

## Statements and Declarations

**Funding** This work was supported by the Estonian Research Council grants PSG728, Tem-TA 120 and PUTJD1252 and the Estonian Centre of Excellence in Artificial Intelligence (EXAI), funded by the Estonian Ministry of Education and Research.

**Author contributions** MV screened and reviewed the articles under analysis in the first round, wrote the first draft and reviewed the articles in the second round; MAK reviewed the articles in the first round, screened and reviewed the articles in the second round, contributed to writing and edited the second draft; KT reviewed and edited the first and second drafts; JA guided the design of the overall approach and reviewed the first and second drafts.

**Conflict of interests** The authors declare no conflict of interest.

**Data availability** No data available

**Appendix 1**

**Preferred Reporting Items for Systematic reviews and Meta-Analyses extension for Scoping Reviews (PRISMA-ScR) Checklist**

| REPORTED ON PAGE # | PRISMA-ScR CHECKLIST ITEM | ITEM | SECTION |
|---|---|---|---|
| **TITLE** | | | |
| 1 | Identify the report as a scoping review. | 1 | Title |
| **ABSTRACT** | | | |
| 1 | Provide a structured summary that includes (as applicable): background, objectives, eligibility criteria, sources of evidence, charting methods, results, and conclusions that relate to the review questions and objectives. | 2 | Structured summary |
| **INTRODUCTION** | | | |
| 2 | Describe the rationale for the review in the context of what is already known. Explain why the review questions/objectives lend themselves to a scoping review approach. | 3 | Rationale |
| 2 | Provide an explicit statement of the questions and objectives being addressed with reference to their key elements (e.g., population or participants, concepts, and context) or other relevant key elements used to conceptualize the review questions and/or objectives. | 4 | Objectives |
| **METHODS** | | | |
| - | Indicate whether a review protocol exists; state if and where it can be accessed (e.g., a Web address); and if available, provide registration information, including the registration number. | 5 | Protocol and registration |
| 3 | Specify characteristics of the sources of evidence used as eligibility criteria (e.g., years considered, language, and publication status), and provide a rationale. | 6 | Eligibility criteria |
| 3 | Describe all information sources in the search (e.g., databases with dates of coverage and contact with authors to identify additional sources), as well as the date the most recent search was executed. | 7 | Information sources* |
| 3 | Present the full electronic search strategy for at least 1 database, including any limits used, such that it could be repeated. | 8 | Search |
| 3-4 | State the process for selecting sources of evidence (i.e., screening and eligibility) included in the scoping review. | 9 | Selection of sources of evidence† |
| 3-4 | Describe the methods of charting data from the included sources of evidence (e.g., calibrated forms or forms that have been tested by the team before their use, and whether data charting was done independently or in duplicate) and any processes for obtaining and confirming data from investigators. | 10 | Data charting process‡ |
| 4-5 | List and define all variables for which data were sought and any assumptions and simplifications made. | 11 | Data items |
| - | If done, provide a rationale for conducting a critical appraisal of included sources of evidence; describe the methods used and how this information was used in any data synthesis (if appropriate). | 12 | Critical appraisal of individual sources of evidence§ |
| 4-5 | Describe the methods of handling and summarizing the data that were charted. | 13 | Synthesis of results |
| **RESULTS** | | | |
| 4 | Give numbers of sources of evidence screened, assessed | 14 | Selection of sources of |
| | for eligibility, and included in the review, with reasons for exclusions at each stage, ideally using a flow diagram. | | evidence |
| 6-8 | For each source of evidence, present characteristics for which data were charted and provide the citations. | 15 | Characteristics of sources of evidence |
| - | If done, present data on critical appraisal of included sources of evidence (see item 12). | 16 | Critical appraisal within sources of evidence |
| 6-8 | For each included source of evidence, present the relevant data that were charted that relate to the review questions and objectives. | 17 | Results of individual sources of evidence |
| 9 | Summarize and/or present the charting results as they relate to the review questions and objectives. | 18 | Synthesis of results |
| **DISCUSSION** | | | |
| 9-12 | Summarize the main results (including an overview of concepts, themes, and types of evidence available), link to the review questions and objectives, and consider the relevance to key groups. | 19 | Summary of evidence |
| 12-13 | Discuss the limitations of the scoping review process. | 20 | Limitations |
| 13 | Provide a general interpretation of the results with respect to the review questions and objectives, as well as potential implications and/or next steps. | 21 | Conclusions |
| **FUNDING** | | | |
| 13 | Describe sources of funding for the included sources of evidence, as well as sources of funding for the scoping review. Describe the role of the funders of the scoping review. | 22 | Funding |

JBI = Joanna Briggs Institute; PRISMA-ScR = Preferred Reporting Items for Systematic reviews and Meta-Analyses extension for Scoping Reviews.

* Where *sources of evidence* (see second footnote) are compiled from, such as bibliographic databases, social media platforms, and Web sites.

† A more inclusive/heterogeneous term used to account for the different types of evidence or data sources (e.g., quantitative and/or qualitative research, expert opinion, and policy documents) that may be eligible in a scoping review as opposed to only studies. This is not to be confused with *information sources* (see first footnote).

‡ The frameworks by Arksey and O'Malley (6) and Levac and colleagues (7) and the JBI guidance (4, 5) refer to the process of data extraction in a scoping review as data charting.

§ The process of systematically examining research evidence to assess its validity, results, and relevance before using it to inform a decision. This term is used for items 12 and 19 instead of "risk of bias" (which is more applicable to systematic reviews of interventions) to include and acknowledge the various sources of evidence that may be used in a scoping review (e.g., quantitative and/or qualitative research, expert opinion, and policy document).

*From:* Tricco AC, Lillie E, Zarin W, O'Brien KK, Colquhoun H, Levac D, et al. PRISMA Extension for Scoping Reviews (PRISMAScR): Checklist and Explanation. *Ann Intern Med. 2018*;169:467–473. doi: 10.7326/M18-0850.

**Appendix 2**

Full search queries used.

PubMed: ("journal article"[Publication Type]) AND ((immersive virtual reality[Title]) AND (randomized controlled trial[Title])) NOT (((non-immersive[Title]) OR review[Title]) OR (meta-analysis[Title]) OR (protocol[Title]) OR (distraction[Title]) OR (training[Title]) OR (physiotherapy[Title]) OR (balance[Title]) OR (stroke[Title]))

Scopus: DOCTYPE(ar) AND TITLE("immersive virtual reality") AND TITLE("randomized controlled trial") AND NOT (TITLE("non-immersive") OR TITLE(review) OR TITLE("meta-analysis") OR TITLE(protocol) OR TITLE(distraction) OR TITLE(training) OR TITLE(physiotherapy) OR TITLE(balance) OR TITLE(stroke))

Web of Science: (DT==("Article")) AND (TI=("immersive virtual reality") AND TI=("randomized controlled trial")) NOT (TI=("non-immersive") OR TI=("review") OR TI=("meta-analysis") OR TI=("protocol") OR TI=("distraction") OR TI=("training") OR TI=("physiotherapy") OR TI=("balance") OR TI=("stroke"))

PsycINFO: PZ("journal article") AND TI ("immersive virtual reality" AND "randomized controlled trial") NOT TI (non-immersive OR review OR protocol OR distraction OR training OR physiotherapy OR parkinson OR balance OR stroke OR meta-analysis)